\documentclass[USenglish,twocolumn]{article}

\ifx\directlua\undefined\ifx\XeTeXcharclass\undefined
  \usepackage[utf8]{inputenc}                           
  \else\RequirePackage[no-math]{fontspec}[2017/03/31]\fi 
  \else\RequirePackage[no-math]{fontspec}[2017/03/31]\fi 
\usepackage[sort&compress,square,numbers]{natbib}
\usepackage{graphicx}
\usepackage{booktabs, makecell}
\usepackage{xr}
\usepackage{amsmath}
\usepackage{hyphenat}
\usepackage{authblk}
\usepackage{lineno}

\newcommand*{\addFileDependency}[1]{
\typeout{(#1)}
\@addtofilelist{#1}
\IfFileExists{#1}{}{\typeout{No file #1.}}
}\makeatother

\title{Electrically tunable plasmonic metasurface as a matrix of nanoantennas}

\author[1,2,*]{LUIS ANGEL MAYORAL-ASTORGA}
\author[1,2]{MASOUD SHABANINEZHAD}
\author[1,2]{HOWARD NORTHFIELD}
\author[1,2]{SPYRIDON NTAIS}
\author[1,4]{SABAA RASHID}
\author[1,4]{EWA LISICKA-SKRZEK}
\author[5]{HAMID MEHRVAR}
\author[5]{ERIC BERNIER}
\author[5]{DOMINIC GOODWILL}
\author[2,3]{LORA RAMUNNO}
\author[1,2,3]{PIERRE BERINI}

\affil[1]{School of Electrical Engineering and Computer Science, University of Ottawa, Ottawa, ON K1N 6N5, Canada}
\affil[2]{NEXQT Institute, University of Ottawa, 25 Templeton Street, Ottawa, ON K1N 6N5, Canada}
\affil[3]{Department of Physics, University of Ottawa, Ottawa, ON K1N 6N5, Canada}
\affil[4]{Canadian Centre for Electron Microscopy, McMaster University, Hamilton, Canada}
\affil[5]{Huawei Technologies Canada, Ottawa, Canada}

\begin{document}

\maketitle

\begin{abstract}
We report the fabrication and characterization of a plasmonic metasurface comprising electrically-contacted sub-wavelength gold dipole nanoantennas, conformally coated by a thin hafnia film, an indium tin oxide layer and a backside mirror, forming metal-oxide-semiconductor (MOS) capacitors, for use as an electrically-tunable reflectarray or metasurface. By voltage biasing the nanoantennas through metallic connectors and leveraging the carrier refraction effect in the MOS capacitors, our measurements demonstrate phase control in reflection over a range of about 30 degrees, with a constant magnitude of reflection coefficient of ~0.5, and the absence of secondary lobes. Comprehensive electromagnetic and quantum carrier models of the structure are developed and are in excellent agreement with the measurements. The metasurface holds promise for use as an optical phased array. 
\end{abstract}


\section{Introduction} 
Electrically-tunable metasurfaces hold significant promise for numerous applications, including dynamic beam forming and focusing, beam steering, and, generally, wavefront shaping \cite{Shalaginov2020,Nemati2018,Shaltout2019}.

Beam steering, for example, is key to LIDAR scanners for self-driving cars and autonomous machines, but is typically achieved using mechanical means, leading to bulky and slow systems which precludes their use in cases that require a high refresh rate. Nanophotonic approaches hold promise for LIDAR applications \cite{Kim2021},

particularly to implement beam steering devices based on optical phased arrays\cite{Heck2017,Berini2022}. Compared to mechanical scanning, optical phased arrays exploiting electronic tuning can enable significantly faster scanning speeds.

An ideal beam steering metasurface should possess several key characteristics \cite{Berini2022}. It should feature a large aperture, allowing for the steering of light across a wide range of angles, without generating diffraction orders. Simultaneously, it should maintain the beam at a constant magnitude with low loss throughout the steering process. In addition, such a surface should be reliable, compact, cost-effective, and solid-state, eliminating the need for mechanical components. It should also offer rapid scanning capabilities to enable swift operations. Optical phased arrays for beam scanning can be implemented using tunable metasurfaces \cite{Berini2022}. Such structures typically consist of a planar arrangement of phase-tunable coherent light emitters, where an individual emitter is referred to as a pixel. A pixel can take various forms and be arranged to form a metasurface that operates in reflection (reflectarray) or in transmission (transmitarray). 

Optical phased arrays based on plasmonic nanoantennas hold promise for meeting the requirements of optical beamsteering. Resonant plasmonic nanoantennas support strongly enhanced fields which confer a high sensitivity to local perturbations in refractive index. They are also nanoscale structures, enabling pixels to be dimensioned to sub-wavelength scales which precludes the emergence of grating (secondary) lobes. Moreover, metal nanoantennas can also serve as device electrodes, ensuring strong optical overlap with the active region. 

Integrating plasmonic nanostructures with a semiconductor enables exploitation of the carrier refraction effect through electrical biasing in an appropriate device configuration. Altering the carrier density by applying a gate voltage provides a pathway to dynamically and actively tune the phase, amplitude and polarization of the emerging light \cite{Calalesina2021, Decker2013, Olivieri2015, Thyagarajan2017, Berini2022}. A promising approach to dynamically tune a metasurface for beam steering involves manipulating the optical properties of a semiconductor used in its construction. Transparent conductive oxides (TCOs), particularly Indium Tin Oxide (ITO), are of strong interest as this material offers high transparency in the visible and near-infrared spectrum, coupled with excellent electrical conductivity, making it an ideal candidate for applications in photonics and optoelectronics. 

Implementing devices that replicate the function of a turning mirror to control a beam generated by a nearby source involves the pixelation and application of a controllable phase gradient across a reflective area. Such a device, termed a tunable reflectarray, allows for steering a monochromatic beam to any desired angle of reflection. To achieve this, it is desirable that the phase of each pixel over the beam cross-section to be continuously variable from $0$ to $2\pi$ without affecting the incident light’s intensity.

The versatility of the tunable reflectarray concept makes it a powerful tool. Much of the ongoing research in beam steering devices, particularly those involving optical metasurfaces or plasmonic nanostructures, is focused on the development of reflectarrays constructed with tunable pixels, incorporating ITO as electro-optical active material.

Huang \emph{et al.} \cite{Huang2016} achieved a phase change of 184° at a wavelength of 1573 nm for a single-gated device comprised of 50 nm thick gold strips on 5 nm of Al$_2$O$_3$, on 20 nm of ITO, on 80 nm of Au. However, their structure displayed a strong dependence of reflectance on the bias voltage. Park \emph{et al.} \cite{park2017} reported phase tuning of 180° in the reflected field using a device stack comprised of Au, Al$_2$O$_3$, ITO and Au, operating at a wavelength of 6 µm. Double-gated devices were reported where the metasurface elements were controlled with two independent voltages driving two MOS field effect regions, thereby enabling wider phase tunability \cite{Shirmanesh2018}. They employed ITO as the active metasurface material and a composite hafnia/alumina gate dielectric. The device exhibited a continuous phase shift from 0 to 303° at a wavelength of $\lambda$ = 1550 nm, however, it displayed a significant dependence of reflectance on the bias voltage. 

More recently, Kim \emph{et al.} \cite{KimIl2020} introduced a beam steering device composed of a 5 nm thick ITO layer sandwiched between two insulator layers (Al$_2$O$_3$/HfO$_2$). By employing two separate bias controls, they managed to control both the amplitude and the phase from 0 to 360$^0$. They reported a side mode suppression ratio (SMSR) of 2.7 dB. An experimental demonstration of beam steering with a metasurface array operating at 1.3 $\mu$m was also recently presented \cite{KimIl2022}. The metasurface unit cell consisted of a metal–dielectric–oxide structure with indium tin oxide as an active layer, modulated by using top fan-out electrodes. The metasurface array was pixelated in two-dimensions and exhibited a phase change of 137°, allowing for two-dimensional beam steering to an angle of 7.3°. The side mode suppression ratio (SMSR) ranges from -9 to 6 dB.

Here we focus on the development of a reflectarray following a nanoscale pixel design based on a gold dipole nanoantenna that enables phase control in reflection, explored theoretically in previous work \cite{Calalesina2021}. We demonstrate phase tuning with a constant magnitude of reflection and the absence of secondary lobes due to the sub-wavelength dimensions of the pixels. In the forthcoming sections, we describe and discuss the device concept, our fabrication techniques and our test set-up, and we present measurements compared with theoretical results based on electromagnetic and quantum carrier models of the structure. We give brief conclusions in the last section of the paper.

\section{Description and operation of the reflectarray}

\begin{figure}[ht!]
\includegraphics[width=\columnwidth]{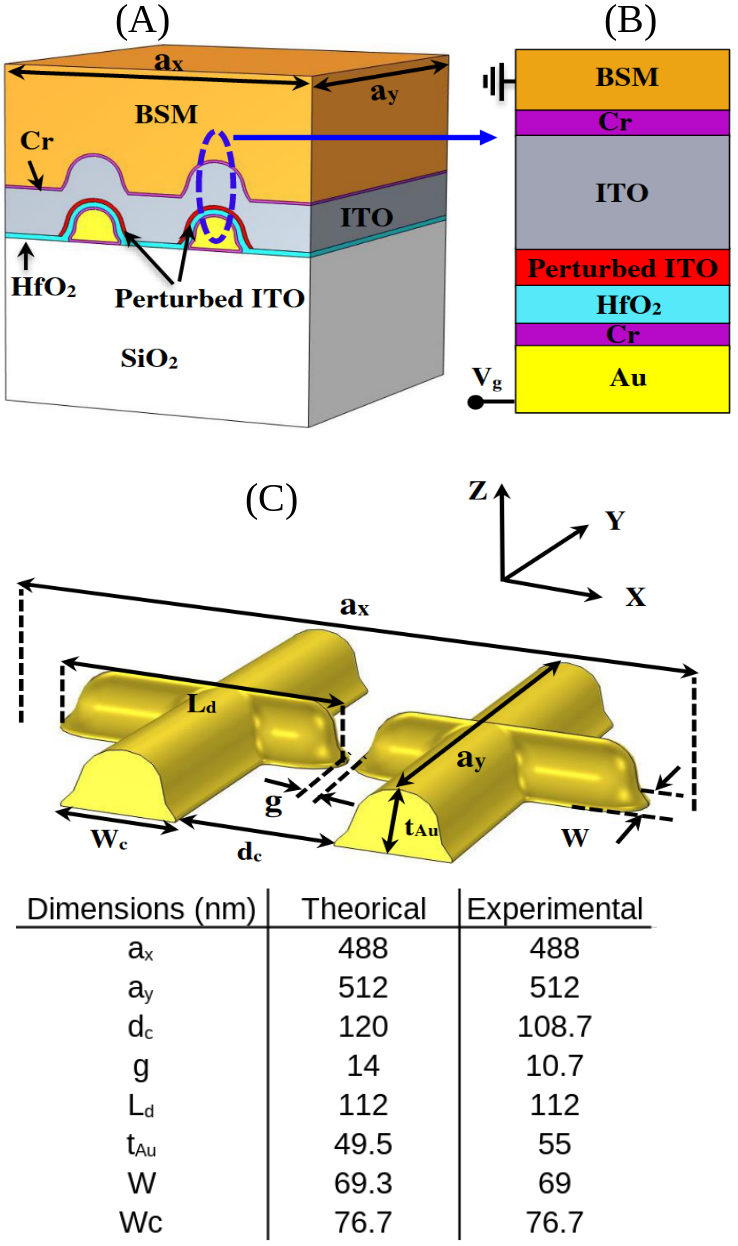}
\caption{Sketch of the sub-wavelength unit cell used in the reflectarray, including the dimensions used in the theoretical model of the experimentally realised structure. (A) Panoramic view of the complete material stack of the unit cell, (B) 1D MOS capacitor enabling electrical control of the optical performance of the pixel, and (C) panoramic view of gold dipole nanoantennas along with connectors.}
 \label{fig:unitcell}
\end{figure}

Figure \ref{fig:unitcell}(A) illustrates the plasmonic unit cell (pixel) of dimensions $a_x$ by $a_y$ employed in our reflectarray. The unit cell exhibits periodicity along both the $x$ and $y$ directions. The pixel design comprises a gold dipole nanoantenna, formed by two rods of length $L_d$, width $W$, and thickness $t_{Au}$. These rods are separated by a gap $g$ and intersected by two perpendicular gold connectors of width $W_c$, positioned at an edge-to-edge distance of $d_c$ from each other, as shown in Fig. \ref{fig:unitcell}(C). The dimensions of the dipole and connectors within the plasmonic unit cell are presented in the table inset in Fig. \ref{fig:unitcell}. 

Within this unit cell, a metal dipole nanoantenna is positioned with its long axis aligned in the $x$-direction. This configuration enables the structure to resonate in its fundamental mode at the desired wavelength when excited by a plane wave or beam that is polarized in the $x$-direction and propagating in the $z$-direction (upward through the SiO$_{2}$ substrate). In the case of this particular polarization (or any polarization in the $z-x$ plane), electrical contacts can be strategically placed along each segment of the antenna in the $y$-direction as shown in Fig. \ref{fig:unitcell}. This arrangement produces a nanoantenna array that can be addressed column-wise when viewing the $x-y$ plane (\emph{cf.} Fig. \ref{fig:actual_dev}(H)). Thus, the structure is designed to serve not only as an antenna but also as device electrodes.

The gold antennas and contacts are supported by a glass (fused silica) substrate. They were deposited on a 3 nm layer of chromium to ensure adherence, as shown in magenta in Figs. \ref{fig:unitcell}(A)-(B). Additionally, a 3 nm layer of chromium was deposited onto the gold antennas to facilitate the subsecuent oxide layer. Following this, the structure is further conformally coated by a 7 nm layer of insulating oxide (hafnia, HfO$_{2}$), followed by a 75 nm thick layer of indium tin oxide (ITO) as the semiconductor, and finally by a ground Ohmic contact designated as the backside metal (BSM) also functioning as an optical mirror, comprised of the metal stack 3 nm Cr + 50 nm Au + 500 nm Cu + 50 nm Pt. This layer arrangement forms electrically a MOS (metal oxide semiconductor) capacitor structure, operating optically as a resonant nanoantenna on an insulator/ITO stack, on a mirror.

The device is constructed such that ITO fills the gap between the antenna arms, which is where the optical fields are enhanced and the structure is most sensitive. When the MOS structure is driven into strong accumulation, a high electron concentration layer is formed within a confined area approximately 4 nm away from the interface of the ITO and the insulator. We refer to the accumulation layer as the perturbed layer in ITO, indicated in red in Fig. \ref{fig:unitcell}(B). Furthermore, ITO undergoes a transition through epsilon-near-zero (ENZ) in this layer as accumulation takes place, resulting in a significant perturbation of the antenna resonance, including its phase. 

The antenna unit cells (pixels), of dimensions $a_x$ by $a_y$, are distributed in a periodic fashion over a designated area measuring $10 \times 10$ $\mu$m$^2$, centered within an electrical fan-out structure to contact an array column-wise. The selection of pixel periodicity is crucial, and is chosen small enough to satisfy the condition $a_x$, $a_y$ $\le\lambda/2$, where $\lambda$ represents the wavelength of operation in the medium through which light is incident on the nanoantennas (glass substrate). By respecting this condition, the arrangement ensures the absence of grating diffraction orders in reflection.

\section{Materials and methods}
\subsection{Fabrication process}
\label{sec:fab}
The fabrication process employed three distinct lithography techniques. A trilayer contact photolithography process was employed to create relatively large features via lift-off, such as a fan-out structure, contact fingers and pads, and the backside metallization. For specific details of this fabrication process, refer to \cite{HowardN2021}. This technique provides better resolution than conventional bilayer photolithography lift-off processes. A traditional single-layer photolithography process was used to create a wet etch mask to define ITO regions. A Helium ion beam lithography technique was utilized to fabricate the smallest features via lift-off, such as the nano-antennas and its crossing contacts to fan-out structures. This technique provides better resolution than e-beam lithography while minimising proximity effects. A detailed description of our Helium ion beam lithography process can be found in \cite{Sabaa2021}.

\begin{figure*}[ht!] 
\centering\includegraphics[width=2\columnwidth]{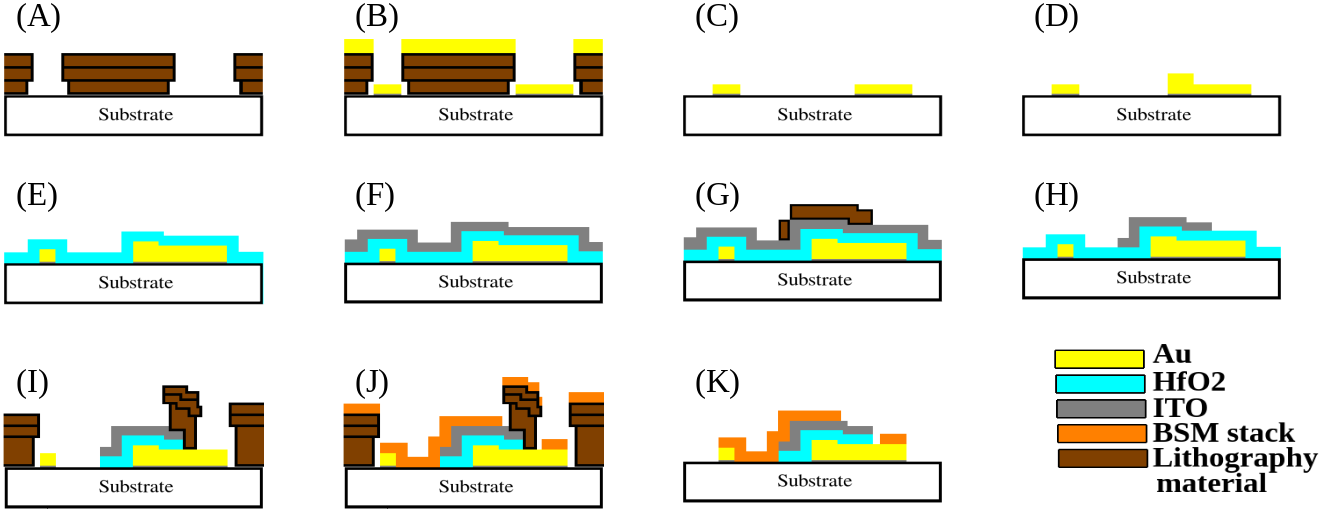}
\caption{Fabrication process flow. (A) Trilayer photolithography stack to define FML on fused silica substrate; (B) FML deposition by evaporation (3 nm Cr + 28 nm Au); (C) Lift-off; (D) Fabrication of nanoantenna arrays by Helium ion beam lithography, metal evaporation, and lift-off (3 nm Cr +  55 nm Au + 3 nm Cr); (E) Atomic layer deposition of 7 nm thick hafnia layer; (F) Sputter deposition of 75 nm thick ITO layer followed by annealing at 350 $^{\text{o}}$C in O$_{2}$; (G) Wet etch mask application; (H) ITO wet etch in bath of TE-100 (Transene); (I) Trilayer photolithography stack to define BSM, followed by removal of exposed hafnia via reactive ion etching (RIE); (J) BSM deposition by evaporation (3 nm Cr + 50 nm Au + 500 nm Cu + 50 nm Pt); (K) Lift-off, revealing completed device.}
\label{fig:fab_flow}
\end{figure*}

The complete fabrication flow is depicted in Fig. \ref{fig:fab_flow}. The fabrication process begins with the patterning of the first metal layer (FML) on a fused silica substrate using a trilayer photolithography process \cite{HowardN2021}, as illustrated in Fig. \ref{fig:fab_flow}(A). The FML was deposited via evaporation as the metal stack 3 nm Cr + 28 nm Au + 2 nm Cr. The top layer of Cr was added to improve adhesion for the subsequent oxide application, as illustrated in Fig. \ref{fig:fab_flow}(A)-(B). It was then lifted-off such that the desired features remained on the substrate, as shown in Fig. \ref{fig:fab_flow}(C). The metal features defined in this step include alignment marks for subsequent process steps and to facilitate experimental characterization, a fan-out structure, contact pads to connect to a nanoantenna array, and various process control structures. Then, the nanoantenna arrays and crossing contacts were fabricated using a Helium ion beam lithography and liftoff process \cite{Sabaa2021}, as the metal stack 3 nm Cr + 55 nm Au + 3 nm Cr, as shown in Fig. \ref{fig:fab_flow}(D). Electron beam lithography was employed to expose alignment marks into the resist enabling Helium ion beam visibility and the overlay of the nanoantennas onto the FML. This was necessary since the FML cannot be imaged through PMMA with a helium ion beam unlike an electron beam. The crossing contacts were aligned and overlapped with the fan out features on the FML to ensure electrical continuity of the nanoantennas to the contact pads. Then we used an atomic layer deposition (ALD) process to conformally deposit a 7 nm thick layer of hafnia, as shown in Fig. \ref{fig:fab_flow}(E). This was followed by the deposition of a 75 nm thick ITO layer via sputtering, which was then annealed at 350 $^{\text{o}}$C in $\text{O}_2$, as shown in Fig. \ref{fig:fab_flow}(F). A wet etch mask was then defined photolithographically, as shown in Fig. \ref{fig:fab_flow}(G), the ITO layer was etched in a bath of TE-100 etchant (Transene Inc.), and the photoresist mask was stripped, as shown in Fig. \ref{fig:fab_flow}(H). The ITO patterns thus formed consisted of disks centered above the nanoantenna arrays and their crossing contacts. Then we defined the BSM using a trilayer photolithograpy process\cite{HowardN2021}, and we removed the exposed hafnia layer using a reactive ion etching (RIE) process prior to BSM deposition, as shown in Fig. \ref{fig:fab_flow}(I). The BSM was then deposited via evaporation as the metal stack 3 nm Cr + 50 nm Au + 500 nm Cu + 50 nm Pt, as shown in Fig. \ref{fig:fab_flow}(J), followed by lift-off to form complete structures, as shown in Fig. \ref{fig:fab_flow}(K). The BSM pattern consisted of circular patches on the ITO disks defining the Ohmic contacts and mirrors thereon, and overlapped metal on the fan-out and pad structures to reduce the resistance and facilitate probing. 

Witness samples were included during the deposition of hafnia and ITO for the purpose of evaluating the films independently. The hafnia layer had a density of 9.61268 g/cm$^3$ and a thickness of 7.76 $\pm$ 0.258 nm, confirmed using X-ray reflectometry (XRR). The electronic properties of the ITO layer were determined using an Ecopia HMS-3000 Hall effect measurement system in Van der Pauw configuration. The optical properties of the deposited ITO and hafnia layers were extracted from ellipsometry measurements using a spectroscopic ellipsometer (UVISEL Plus, Horiba Scientific). See Section \ref{sup-section:ITOandHafniaCha} under supplementary material for details.

\subsection{Electro-optic test set-up and characterization techniques}

\begin{figure}[ht!]
\centering\includegraphics[width=0.9\columnwidth]{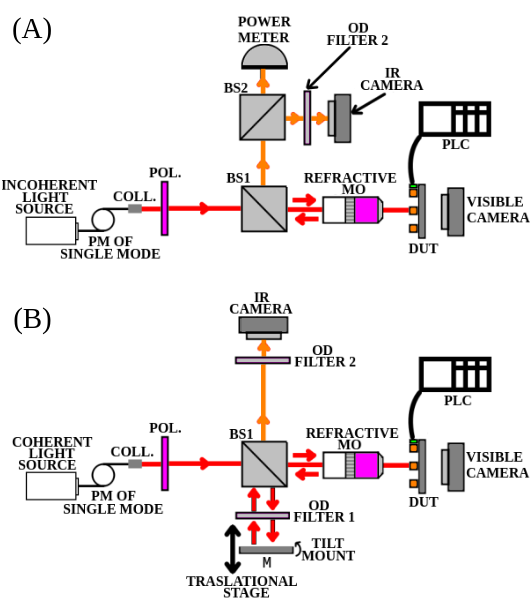}
\caption{Confocal microscope for measuring (A) the magnitude of the reflection coefficient $|\Gamma(V)|$, and (B) its relative phase $\phi(V)$.}
\label{fig:sertup}
\end{figure}

To investigate the reflectarray's electro-optical performance, we used two custom-built confocal microscopes. Fig. \ref{fig:sertup}(A) shows the setup used to measure the magnitude of the reflection coefficient $|\Gamma(V)|$, and Fig. \ref{fig:sertup}(B) shows the setup used to measure the relative phase shift $\phi(V)$, both at normal incidence and at bias voltage $V$. In both setups, The input light beam was delivered through a polarization-maintaining single-mode optical fiber (PMOF) and collimated using a fiber collimation package (F220APC-1550, Thorlabs). A polarizer (LPNIR100, Thorlabs) was inserted to fix the polarization orientation parallel to the nano-antennas. The incident beam was focused onto the reflectarray using a plan apochromatic NIR objective (IR18091161 50$\times$, NA = 0.42, BoliOptics). The reflected beam was redirected by beamsplitter BS1 (BSW29, Thorlabs) to the measurement instruments. For the setup shown in Fig. \ref{fig:sertup}(A), beamsplitter BS2 allowed monitoring the beam spot using an IR camera (Micronviewer 7290A, Electrophysics) while taking power measurements (81525A, Agilent/HP). For the setup shown in Fig. \ref{fig:sertup}(B), beamsplitter BS1 and mirror M produced a reference beam used to interfere with the beam reflected by the reflectarray on the IR camera (this setup effectively integrates a Michelson interferometer at the input of the confocal microscope).

We used two different light sources for our measurements. An incoherent light source was used to measure the magnitude of the reflection coefficient (Fig. \ref{fig:sertup}(A)), whereas a coherent light source was used to measure the relative phase shift (Fig. \ref{fig:sertup}(B)). The incoherent light was delivered by a super-continuum white light source (SuperK Extreme, NKT Photonics) with an acousto-optic tunable filter (SuperK SELECT, NKT Photonics), whereas the coherent light was delivered by a tunable laser (TUNICS-Plus CL, 1510 nm to 1640 nm, Photonetics).

The sub-wavelength reflectarray was designed to reflect the incident beam without secondary lobes (which would be apparent in our set-up on images captured by the IR camera). It is desirable for the incident beam spot to be smaller than the area of the reflectarray (10$\times$10 $\mu$m$^2$) to minimize diffraction from its edges\cite{CalaLesina2020}.

To ensure this, the beam spot diameter was measured for both of our sources (super-continuum and the tunable laser) using the knife-edge method\cite{Araujo2009}, resulting in a 8.5 $\mu$m diameter, fulfilling this requirement.

To apply bias voltages to the device under test (DUT), we used a custom probe card (Accuprobe) to simultaneously contact the 42 pads of a DUT: 2 pads connected in common to the device's ground, and 40 pads connected columnwise to the nanoantenna array to apply the gate voltages. Alignment of the probe card to a DUT was facilitated by a miniaturised microscope camera positioned on the backside of the DUT, as depicted in Fig. \ref{fig:sertup}. We used a programmable logic controller (PLC) (Productivity2000, Automation Direct) with analog output modules (P2-08DA-2, 8 outputs, Automation Direct) and two 16 bit digital to analog converters (P2-16DA-2, 32 outputs, Automation Direct) to apply a bias voltage individually to every pad of the DUT. Each analog output had a voltage resolution of 0.305 V/count, with a maximum possible count of 65536 (16 bits), and a maximum/minimum DC output voltage of $\pm$10 V. The maximum available current per output was 10 mA.

We measured the magnitude of the relative reflection coefficient $|\Gamma(V)|$ at different bias voltages using the setup shown in Fig. \ref{fig:sertup}(A). $|\Gamma(V)|$ was taken as $\sqrt{P_m/P_{sub}}$, where $P_m$ is the power reflected by the reflectarray, and $P_{sub}$ is the power reflected by a nearby region of the substrate without nanoantennas but coated with hafnia, ITO and the backside Au mirror. $P_{sub}$ can be used as a reference power because the hafnia and ITO layers are very thin and the backside mirror is a good approximation to a perfect mirror at the wavelengths of interest. Optical density filter 2 (OD 2) was adjusted to avoid saturating the IR camera. We kept the applied bias voltage equal and constant on all pads while sweeping the wavelength.

The phase shift measurement relied on the intensity change resulting from the interference between the beam reflected by the Device Under Test (DUT) and the reference beam reflected by mirror M (Fig. 3(B)). First we model the overlapping beams as two interfering waves. We write the wave representing the reference beam as $w_1(z,t)=A\cos{(kz-\omega t)}$, and the wave representing the beam reflected by the reflectarray as $w_2(z,t)=|\Gamma(V)|B\cos{(kz-\omega t + \phi (V))}$. Setting $z=0$ as the camera location and $t=0$ yields $w_1=A$ and $w_2=|\Gamma(V)|B\cos{(\phi(V))}$. The intensity of the interference of the beams on the camera can be expressed

\begin{equation}\label{eq:intensity_IV}
    \begin{split}
     I(V)=&[A+|\Gamma(V)|B\cos{(\phi(V))}]^2\\
     =&A^2+2AB|\Gamma(V)|\cos{(\phi(V))}+\\
     &|\Gamma(V)|^2 B^2 \cos^2{(\phi(V))},
    \end{split}
 \end{equation}

Initially, the bias voltage was set to zero. To achieve equal intensity for the reflected and reference beams on the camera, OD 1 and OD 2 were carefully adjusted. This involved individually blocking one beam's optical path while adjusting the Optical Density (OD) filter of the other.

To align both beams on the camera, the tilt mount of mirror M was adjusted until the position of the reference beam almost perfectly overlapped with the reflected beam. Subsequently, the translation stage of mirror M was fine-tuned to achieve total constructive interference. The intensity $I(0)$ was captured under the conditions that the intensity of individual beams $w_1$ and $w_2$ were the same and constructively interfering ($A=|\Gamma(0)| B$ and $\cos{(\phi(0))}=1$). From equation \eqref{eq:intensity_IV} at V = 0 V, we deduced $A=\frac{\sqrt{I(0)}}{2}$

Then a non-zero and identical bias voltage was simultaneously applied to all pads of the device, and $I(V)$ was measured. By assuming that the intensity change in the resulting interference patterns was solely due to the phase shift or that the magnitude of the relative reflection coefficient at a given wavelength is almost constant ($|\Gamma(V)|\approx|\Gamma(0)|$),  we solved $\cos{(\phi(V))}$ in eq. \eqref{eq:intensity_IV} using the quadratic formula:

 \begin{equation}\label{eq:quadraticEqofCosphi}
   \begin{split}
    0=&(A^2-I(V))+2A^2\cos{(\phi(V))} +\\
    &A^2 \cos^2{(\phi(V))}.
    \end{split}
 \end{equation}

The solutions of Eq. \eqref{eq:quadraticEqofCosphi} are:
\begin{align}\label{eq:sols_quadraticCosphi}
\cos{(\phi(V))}=&\frac{-A + \sqrt{I(V)}}{A} \\ 
\cos{(\phi(V))}=&\frac{-A - \sqrt{I(V)}}{A}
\end{align} 
\noindent where the second solution is dropped for producing values out of range of the cosine function. Isolating the phase shift $\phi(V)$ from Eq.  \eqref{eq:sols_quadraticCosphi}, yields:
\begin{equation}\label{eq:relPhase}
    \phi(V)=\cos^{-1}\Big({\frac{\sqrt{I(V)}-A}{A} }\Big),
\end{equation}
where $I(V)$  has been measured and $A$ determined as described above. See section S\ref{sup-section:phase_measurement_example} under supplementary material for a an example calculation of the phase shift from the raw data.

\subsection{Capacitance vs. voltage (C-V) measurements}

We measured individual capacitance-voltage (C-V) characteristics for every nanoantenna column using an HP 428a LCR meter, at an AC voltage of $\pm$30 mV and a frequency of 10 kHz. The DC bias voltage was increased from -2 V in steps of 0.05 V until the breakdown voltage (BDV) was reached in accumulation. The C-V measurements were performed after the electro-optic measurements because the capacitor structures become compromised when the BDV is reached.

From the C-V characteristics, we determined the dielectric constant (DC relative permittivity) of hafnia using: 
\begin{equation}\label{eq:epox}
\epsilon_{ox}=C_{max} \frac{t_{ox}}{A_p\epsilon_0},
\end{equation}

\noindent where $C_{max}$ is the maximum capacitance measured in accumulation, $t_{ox}$ is the thickness of the hafnia layer, $A_p$ is the capacitor area probed by pad $p$, and $\epsilon_0$ is the vacuum permittivity. The breakdown field (BDF) in MV/cm was calculated using 
\begin{equation} \label{eq:BDF}
\text{BDF}=\text{BDV} \times 10/t_{ox,nm},
\end{equation}

\noindent where the breakdown voltage BDV is in V and $t_{ox,nm}$ is the thickness of the hafnia layer in nm.

\subsection{Electrostatic modelling}
\label{sec:static}
Classical models such conventional drift-diffusion (CDD) \cite{Innem2020, Rajput2023} are widely used to compute the voltage-dependent carrier density profile in the semiconductor of MOS capacitors. However, when the MOS structure is incorporated into a plasmonic structure or metasurface, and the semiconductor is an epsilon-near-zero medium (such as ITO), then effects such as quantum confinement, quantum tunneling, quantum compressibility and electron diffraction/interference need to be considered. These factors are crucial to accurately compute the profile of the carrier density \cite{Luscombe1992,Fiegna1997,Sojib2022}. The details of the profile significantly impact the accuracy of the resulting permittivity distribution, particularly regarding the number and location of epsilon-near-zero points \cite{Masoud2022}. Here, we used self-consistent Schr\"odinger-Poisson (SP) coupling to compute the carrier density profile in ITO under accumulation \cite{Masoud2022}. This model is theoretically more rigorous than CDD because it takes the aforementioned quantum effects into account. We used the Semiconductor Physics node available under the Semiconductor Module of COMSOL 6.1, and the Schr\"odinger-Poisson interface with the equilibrium study \cite{comsol6v1}. 

In materials like ITO, which are sensitive to local refractive index perturbations, the SP model shows significant deviations from classical models in predicting carrier density under accumulation. The Conventional Drift-Diffusion (CDD) model \cite{Masoud2022} predicts the highest carrier density at the HfO2-ITO interface, decreasing exponentially from the boundary. This prediction contrasts with the SP model, which suggests a lower density at the interface and a peak within the ITO. The CDD model, treating carriers as a classical free electron gas, overlooks quantum effects such as quantization, confinement, and diffraction/interference, which are accounted for in the SP model. These quantum effects lead to findings like the reduced carrier density at the HfO2-ITO interface due to quantum confinement and oscillatory profiles at higher gate voltages. At sufficiently high voltages, the SP model predicts the carrier density in ITO crossing two ENZ points, contrasting with the single ENZ point by the CDD model. This discrepancy significantly affects the permittivity profile and the optical response of the material \cite{Masoud2022}.

The Poisson and Schr\"odinger equations are coupled second-order nonlinear differential equations, and solving this system is nontrivial, requiring an iterative approach. To start the iterations, we computed the potential distribution throughout the 1D MOS capacitor (Fig. \ref{fig:unitcell}(B)) via the CDD and used it as an input to the SP model. Since the Cr is just 3 nm, we assumed that voltage is applied to the gold and ignored the Cr in electrostatic simulation. To obtain the potential distribution, we applied a gate voltage to the metal and grounded the ITO. The metal work function was set to 5.1 eV, whereas the electron affinity, bandgap energy, static relative permittivity, and doping concentration of the ITO were set to 4.8 eV, 2.8 eV, 9.1, and 2.65 $\times 10^{20}$ cm$^{-3}$, respectively.

The potential distribution obtained from the CDD was then input into the SP model by updating the potential term therein. In the next step, the carrier density profile $N(d)$ is computed using the sum of statistically weighted densities of the eigenfunctions obtained from the updated Schr\"odinger equation via \cite{Luscombe1992, Masoud2022}
\begin{equation} \label{eq:theo_n}
 N(d)=\sum_i W_i|\phi_i(d)|^2, 
\end{equation}
where $W_i$ represents the weight factors, $\phi_i$ is the $i^{th}$ normalized eigenfunction associated with eigenenergy $E_i$, and $d$ denotes the distance from the oxide surface. Then, the new potential distribution is computed by solving Poisson's equation using the updated charge distribution $N(d)$ (Eq. \eqref{eq:theo_n}) in the charge density term therein. This procedure is repeated until the electrostatic potential distribution converges. A uniform mesh of step size 0.1 nm was used to discretize the structure. Further details on setting up the problem in COMSOL can be found elsewhere \cite{Masoud2022}.

We then used the electron density distribution computed by the SP model $N(d)$ in the Drude equation \cite{Kittel2004} to obtain the spatially-dependent relative permittivity in the perturbed ITO:
\begin{equation}
\label{eq:drude_masoud1}
 \epsilon(d,\omega)=\epsilon_\infty - \frac{\omega_D(d)^2}{\omega^2+i\gamma_D\omega},
\end{equation}
where $\epsilon_\infty$ is the high-frequency relative permittivity and $\gamma_D$ is the collision frequency, which were taken as 3.92 and $4.4 \times 10^{13}$ rad/sec, respectively \cite{Masoud2022}. $\omega_D(d)$ is the spatially-dependent plasma frequency of the ITO obtained via
\begin{equation} \label{eq:drude_masoud}
 \omega_D(d)=\sqrt{\frac{N(d)}{n_b}}\omega_{D,b}
\end{equation}
where $N(d)$ is obtained for an applied bias voltage of interest. $\omega_{D,b}$ is the bulk plasma frequency of ITO and was fixed to $1.55 \times 10^{15}$ rad/sec \cite{Masoud2022}.

\subsection{Optical modelling}
To investigate the gate-tunable optical response of the reflectarray, we used the relative permittivity distribution in ITO computed for different applied gate voltages as described in Subsection \ref{sec:static} as well as the experimentally-determined dimensions of the nanoantenna array (Fig. \ref{fig:unitcell}). The carrier density in the perturbed layer is a function of distance from the hafnia-ITO interface, so we projected the 1D profile computed for the 1D MOS (Fig. \ref{fig:unitcell}(B), red region) over the 3D unit cell (Fig. \ref{fig:unitcell}(A), red region).

We used the Electromagnetic Waves, Frequency Domain interface of the Wave Optics module in COMSOL 6.1 for the optical modelling \cite{comsol6v1}, and we computed the complex reflection coefficient of a unit cell \emph{vs.} wavelength at various applied bias voltages. The wave equations derived from Maxwell's equations were solved by applying material properties and boundary conditions \cite{Berini2000}:

\begin{align}
 \mathbf{\nabla}\times\mathbf{\nabla}\times \mathbf{E}(x,y,z)-k^2\epsilon_r(x,y,z)\mu_r\mathbf{E}(x,y,z)&=0 \\ 
 \mathbf{\nabla}\times\big ( \frac{1}{\epsilon_r(x,y,z)}\mathbf{\nabla} \times \mathbf{H}(x,y,z) \big)-k^2\mu_r\mathbf{H}(x,y,z)&=0 ,
\end{align}
where $\mu_r$ is the relative permeability (set to 1), $\epsilon_r(x,y,z)$ is the relative permittivity distribution describing the structure, $k$ is the free-space wavenumber, and $\mathbf{E}$ and $\mathbf{H}$ are the electric and magnetic field vectors.

The structure varies along the $z$-axis and is periodic along the $x$ and $y$ directions (Fig. \ref{fig:unitcell}). The propagation and electric field directions of the incident light were taken along the $+z$ and $x$ directions, respectively. Periodic boundary conditions were applied on the $x$ and $y$ boundaries, and a source port was placed along the bottom to excite the structure and record the reflected wave (the excitation propagates through the SiO$_2$ substrate, in accord with the experimental situation). The modeling domain in the $z$ direction was truncated and two perfectly matched layers (PMLs) were added to the top and bottom of the unit cell to absorb the incident waves. 

In our optical model, the refractive index of hafnia was taken from experimental ellipsometry data (see section S\ref{sup-section:ITOandHafniaCha}), the relative permittivity of Au and Cr were taken from Johnson and Christy \cite{Johnson1972} and Raki\'c \emph{et al.} \cite{Rakic98}, respectively, and the voltage-dependent relative permittivity distribution of the ITO - obtained from our electrostatic calculations in Subsection \ref{sub:staticPerformance} - was used. We used mesh mapping along the top boundaries of the perturbed region in ITO and swept it throughout the 3D domain of this region in a step of 0.1 nm. A tetrahedral mesh was used for the remaining domains.

\section{Results and discussion}
\subsection{Fabrication}
\begin{figure*}[ht!] 
\centering\includegraphics[width=\textwidth]{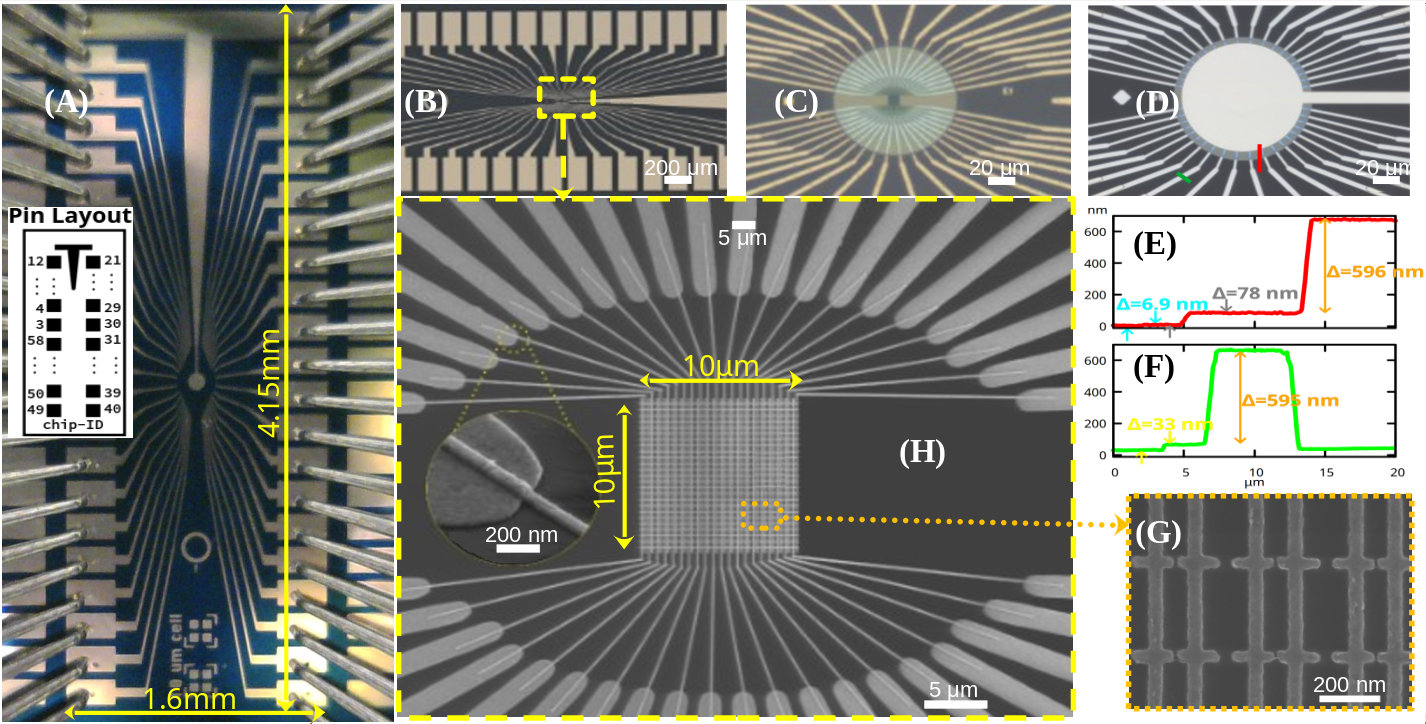}
\caption{Images of fabricated devices. (A) Microscope image of complete device with pins landed on contact pads. Inset: Pad layout of the device and notation. The same layout was used for the pins of the probe card. (B) Microscope image (50$\times$) of the first metal layer (FML) and the nanoantenna array. (C) Microscope image (500$\times$) of the device center showing the fan-out fingers reaching beneath the circular ITO patch. (D) Microscope image (500$\times$) of the device center after backside metal deposition. (E) and (F) AFM scans along two locations showing step changes in the material stack due to hafnia, ITO and backside metal (scan taken along red line sketched in Part (D)), and step changes in the material stack due to the first metal layer and and the backside metal layer (scan taken along green line sketched in Part (D)). (H) HIM image at low magnification revealing the nanoantenna array and the nanoscale fan-out fingers connecting the array to the large-scale fan-out fingers on the FML (high-magnification inset shows good step coverage of these fingers). (G) HIM image at high magnification of a few unit cells of the nanoantenna array.}
\label{fig:actual_dev}
\end{figure*}

Completed devices produced by the fabrication process described in Subsection \ref{sec:fab} are shown in Fig. \ref{fig:actual_dev}, highlighting various stages in the fabrication flow. Fig. \ref{fig:actual_dev}(A) shows a completed device being probed, and the inset depicts the pad, pin layout and numbering. Fig. \ref{fig:actual_dev}(B) shows the device after FML formation and fabrication of the nanoantennas in the central area of the device. Fig. \ref{fig:actual_dev}(C) shows the central area of the device after hafnia and ITO deposition/lift-off (central circular patch). In Fig. \ref{fig:actual_dev}(D), the same area is shown after BSM formation, which is the final step in the fabrication flow. Figs. \ref{fig:actual_dev}(E) and \ref{fig:actual_dev}(f) present AFM scans taken along the red and green lines sketched on Fig. \ref{fig:actual_dev}(D). Fig. \ref{fig:actual_dev}(E) shows step changes in the material stack corresponding to the 6.9 nm thick hafnia layer, the 78 nm thick ITO patch, and the 596 nm thick BSM, whereas Fig. \ref{fig:actual_dev}(F) shows the step changes in the material stack corresponding to the 33 nm thick FLM and 595 nm thick BSM. In Fig. \ref{fig:actual_dev}(H), a low-magnification helium ion microscope (HIM) image reveals the nanoscale fan-out structure connecting the columns of the nanoantenna array to the large-scale fan-out structure on the FML. The high-magnification inset shows good step coverage of these fingers. Fig. \ref{fig:actual_dev}(G) is gives a high-magnification HIM image of a nanoantenna array, from which feature dimensions are measured and reported in the table inset to Fig. \ref{fig:unitcell}.

\subsection{Electrostatic performance} \label{sub:staticPerformance}
\begin{figure}[htbp]
\includegraphics[width=\columnwidth]{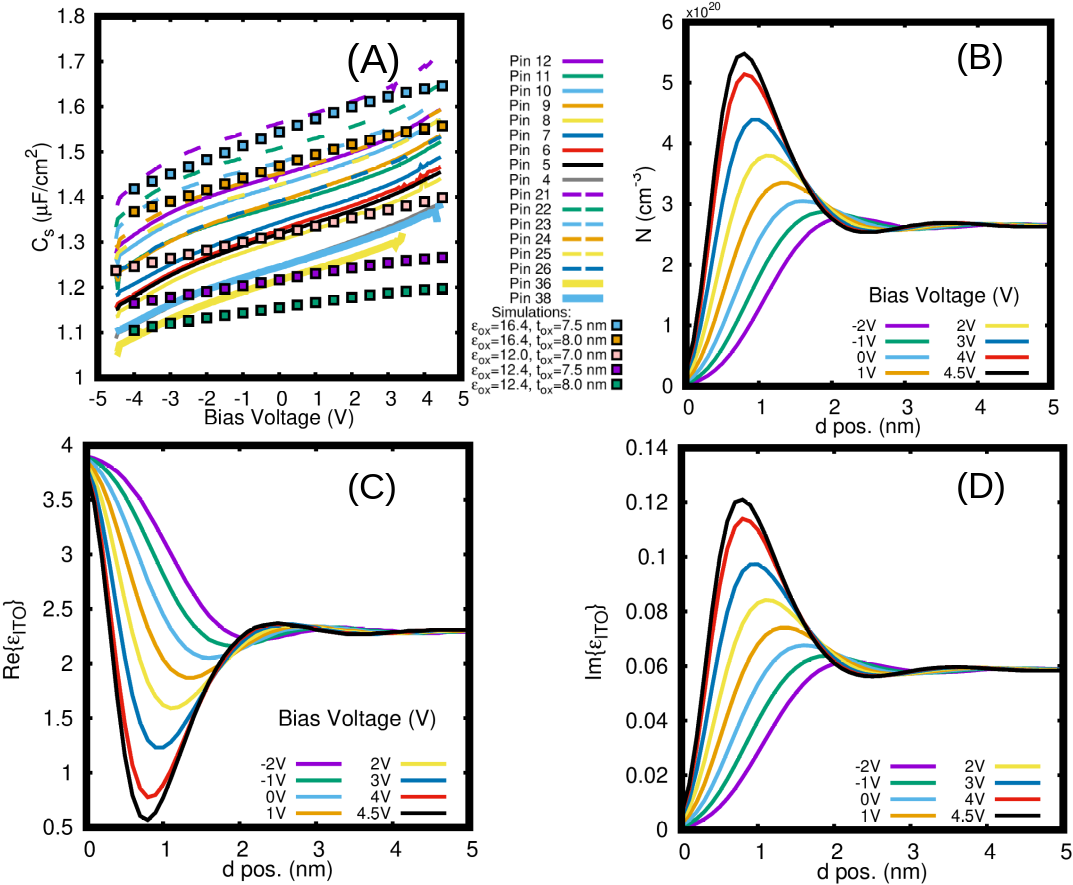}
\caption{Electrostatic performance of sub-wavelength reflectarray. (A) Capacitance per unit area \emph{vs.} applied voltage (C-V curves), measured for several device columns, identified by pad and pin number. The computed C-V curve is also shown for comparison. (B) Computed carrier density profile from the hafnia-ITO interface into ITO for different bias voltages. (C) and (D) Corresponding computed real and imaginary parts of the relative permittivity distribution into ITO.}
\label{fig:calculations1}
\end{figure}

In Figure \ref{fig:calculations1}(A), the graph depicts the computed capacitance per unit area plotted against the voltage applied to the sub-wavelength reflectarray, with the data point represented by colored squares. The computations involved three different values for the DC relative permittivity of hafnia, $\epsilon_{ox}=12$, $12.4$ and $16.4$, combined with three thickness values, $t_{ox} = 7$, $7.5$ and $8$ nm. These values were chosen to encompass the measured experimental C-V characteristics, as discussed subsequently in Section \ref{sec:Hafnia}. Specifically, the combination $\epsilon_{ox}=12$ and $t_{ox} = 7$ nm generates theoretical C-V characteristics that are representative of the average experimental measurements. The C-V theoretical result was obtained by taking the derivative of the charge per unit area in ITO relative to the applied voltage, the former computed from the associated carrier density profiles. The corresponding measured C-V characteristics plotted on Fig. \ref{fig:calculations1}(A) will be discussed in Subsection \ref{sec:Hafnia}.

Figure \ref{fig:calculations1}(B) shows the electron carrier density profiles for $\epsilon_{ox}=12$ and $t_{ox}=7$ nm. The electron carrier density was calculated inside the perturbed region \emph{vs.} distance from the hafnia-ITO interface for the gate voltages listed in the legend, computed using the SP model. It is clear that the perturbation extends to about 4 nm into the ITO from the hafnia-ITO interface ($z=0$). For each applied voltage, a reduction in the carrier density is noted at the hafnia-ITO interface due to quantization of the electron states thereon. At gate voltages from -2 to 0 V the perturbed region is in depletion. By increasing the gate voltage, the carrier density increases, and the region transitions from depletion to accumulation at 1V. As noted in Fig. \ref{fig:calculations1}(B), for positive voltages, the lower carrier density at the interface is followed by a clear maximum inside ITO, which grows and shifts toward the hafnia-ITO interface with increasing voltage and electron accumulation. The observed oscillatory behavior in the carrier density profile is due to electron wavefunction diffraction and interference \cite{Lee2013}.

Figures \ref{fig:calculations1}(C) and \ref{fig:calculations1}(D) show the real and imaginary parts of the relative permittivity of ITO, respectively, computed by inserting the electron carrier density profile shown in Fig. \ref{fig:calculations1}(B) into the Drude equation (eqs. \eqref{eq:drude_masoud1} and \eqref{eq:drude_masoud}), at the free-space operating wavelength of $\lambda=1560$ nm. As noted in Figs. \ref{fig:calculations1}(C) and \ref{fig:calculations1}(D), the real and imaginary parts of the permittivity approach 4 and 0, respectively, at the hafnia-ITO interface for all gate voltages, due to electron depletion that develops thereon (Fig. \ref{fig:calculations1}(B)), driven by the quantization of the electron states. A dip (bump) is then clearly noted in the real (imaginary) part of the permittivity at bias voltages $\ge 0$ V. This dip (bump) occurs at the maximum
of the carrier density in ITO (Fig. \ref{fig:calculations1}(B)) and reduces (increases) with increasing gate voltage. The electric fields become localised and enhanced in regions where the relative permittivity of ITO becomes small \cite{Masoud2022}.

The permittivity of ITO experiences perturbation through the carrier refraction effect induced in the material (recall that the ITO is the semiconductor in MOS capacitors). Voltage gating facilitated by metallic connectors induces a carrier density perturbation over a thin ITO region near the nanoantenna under accumulation and depletion. During accumulation, the carrier density increases, leading to a blue-shift of the epsilon-near-zero wavelength ($\lambda_{ENZ}$) in the perturbed ITO region. The local changes in carrier density within the thin ITO layer, induced by accumulation or depletion in the MOS capacitor, lead to a local variation in permittivity due to the carrier refraction effect. As this perturbed ITO layer is situated in the enhanced field region near the plasmonic nanostructure, its refractive index variation alters the resonance of the nanoantenna, thereby influencing the reflection properties of the reflectarray, particularly its phase shift.

\subsection{Electro-optical performance}\label{sub:electroOpticalPerformance}
\begin{figure}[htbp]
\includegraphics[width=\columnwidth]{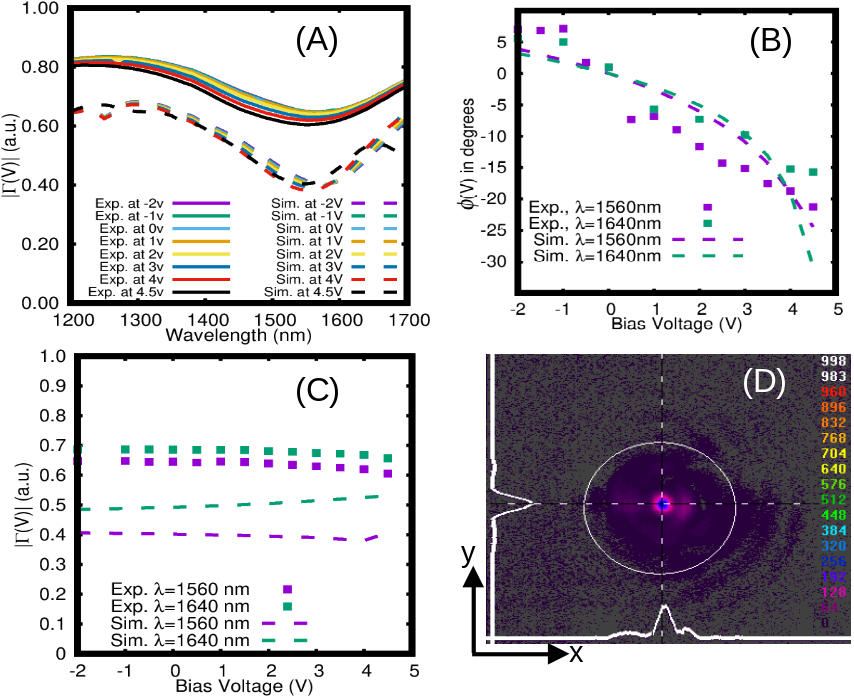}
\caption{Electro-optical performance of sub-wavelength reflectarray. (A) Measured and computed $|\Gamma|$ response for bias voltages between -2 and 4.5 V. (B) Measured and computed relative phase shift $\phi$ \emph{vs.} bias voltage at two operating wavelengths. (C) Measured and computed $|\Gamma|$ \emph{vs.} bias voltage at two operating wavelengths. (D) Intensity contours of beam reflected from sub-wavelength reflectarray, captured by IR camera at zero bias voltage. Horizontal and vertical cross-sectional profiles are shown as the white plots.}
\label{fig:measurements}
\end{figure}

Theoretical and experimental results summarising the electro-optical performance of the sub-wavelength reflectarray are presented in Fig. \ref{fig:measurements}. In Fig. \ref{fig:measurements}(A), the measured wavelength response of $|\Gamma|$ exhibits a prominent dip at 1560 nm, indicating the resonance wavelength for strong coupling of the incident beam to the nanoantennas. Notably, $|\Gamma|$ shows minimal variation across different applied bias voltages. 

The experimental resonance wavelength closely matches the simulation results, thus validating the accuracy of the model. However, it is worth noting that the experimental $|\Gamma|$ are slightly larger than the calculations. The difference between them is 0.15 off-resonance and 0.24 on-resonance. This difference is due to background light present during the measurements and imperfect coupling due to slight misalignment of the beam to the reflectarray.

Figure \ref{fig:measurements}(B) gives the relative phase shift as a function of bias voltage for two  wavelengths of operation, referenced to the phase shift at 0 V, calculated using eq. \eqref{eq:relPhase}. The wavelengths of interest are the resonant wavelength (1560 nm) and the maximum wavelength accessible on our tunable laser (1640 nm). On resonance ($\lambda=1560$ nm), the relative phase shift was measured as $7.4^\text{o}$ at -2 V and $-22.8^\text{o}$ at 4.5 V, resulting in a voltage-tuned phase range of $30^\text{o}$. Conversely, off-resonance ($\lambda=1640$ nm), the voltage-tuned phase range is smaller, spanning $28.6^\text{o}$ starting from $9.1^\text{o}$ at -2 V and reaching $-19.5^\text{o}$ at 4.5 V.

Figure \ref{fig:measurements}(C) gives the magnitude of the relative reflection coefficient as a function of bias voltage at the two selected wavelengths. The magnitude of the reflection coefficient remains flat over the bias voltage range investigated. Specifically, at the wavelength of 1560 nm, the experimental $|\Gamma|$ exhibits a flatness of 4.9\%, about an average of 0.63. In comparison, the theoretical flatness is 4.4\%, about an average of 0.4. Likewise, at the wavelength of 1640 nm, the experimental $|\Gamma|$ has a flatness of 3.3\% about an average of 0.68, whereas the theoretical flatness is 5.2\% about an average of 0.5. The flatness is defined as the ratio of the maximum deviation to the average value, multiplied by 100. 

The insertion loss (IL) in dB of the reflectarray is defined as IL = $-20 \log_{10}|\Gamma|$. The experimental IL at 1560 nm is 4 dB, and at 1640 nm it is 3.35 dB. In contrast, the theoretical IL values are higher, 8 dB at 1560 nm and 6 dB at 1640 nm. The difference between them is attributed to background light present during the measurements and imperfect coupling due to slight misalignment of the beam to the reflectarray, as mentioned previously.

Figure \ref{fig:measurements}(D) provides a visualization of the reflected beam spot from the reflectarray under no bias, including horizontal and vertical cross-sections along the center of the beam spot. Notably, the spot is very clean and absent of diffraction lobes. The vertical cross-section displays a well-formed Gaussian distribution, whereas the horizontal cross-section exhibits a weak diffraction pattern caused by scattering effects from the vertical edges of the reflectarray. The reflected beam maintained the same appearance as the bias voltage was varied over the range of Figure \ref{fig:measurements}(C), including the weak diffraction pattern, owing to the flatness of the magnitude of the reflection coefficient.

These scattering effects are caused by the incident polarization and boundary conditions acting on the vertical edges of the reflectarray. Scattering from metallic edges occurs differently depending on whether the polarisation is hard or soft - the polarization is said to be hard when the boundary condition that applies is a Neumann condition, and soft for a Dirichlet condition \cite{Banalis1989}. The scattering pattern from a metallic edge is broader for the hard polarization than for the soft polarization, yielding an apparent larger beam diameter upon reflection. The incident light beam is linearly polarized and oriented horizontally, so $E_x\neq 0$ and $E_y=0$. As a consequence the electric field across the top and bottom (horizontal) edges of the reflectarray corresponds to the soft polarization, but across the right and left (vertical) edges it corresponds to the hard polarization yielding a broad scattering pattern along the $x$-axis. This undesired effect of diffraction from edges has been previously reported in other experimental work \cite{Kenta2019,Alshehab2019}.

\subsection{Electrical characteristics of the hafnia layer} \label{sec:Hafnia}
Figure \ref{fig:calculations1}(A) illustrates the voltage-dependent specific capacitance (C-V characteristics) measured for a reflectarray. The measurements exhibit repeatability and reveal distinct accumulation patterns in the capacitors of each column, identified by pin number. The raw capacitance measurements were offset on average by $0.0781$ pF relative to the calibrated probes, leading to an overestimation of the specific capacitance by $\Delta C_s = 0.1$ $\mu F/cm^2$. This offset was subtracted from the data presented in Figure \ref{fig:calculations1}(A). More details on this point are given in Section S\ref{sup-section:rawdata} of the supplementary document. The resulting C-V characteristics highlight variations in specific capacitance among the columns, suggesting discrepancies that should ideally converge into a single curve. These deviations are attributed to variances in the hafnia layer.

Analysis of the measured C-V data allows the dielectric constant (DC relative permittivity) of the hafnia layer to be extracted using Eq. \eqref{eq:epox}, assuming a constant hafnia thickness of $t_{ox}=7.76 \pm 0.258$ nm (\emph{cf.} Subsection \ref{sec:fab}, XRR measurements). The dielectric constant was calculated for each pin and ranged between $\epsilon_{ox}=15.85 \pm 0.511$ for pin 21 and $\epsilon_{ox}=12.43 \pm 0.401$ for pin 38, corresponding to the maximum and minimum capacitance values shown in Fig. \ref{fig:calculations1}(A). These pins are positioned at the top and bottom of the device, respectively, indicating a position-dependent variation in the permittivity of the hafnia layer across the device.

The dielectric strength of the Hafnia layer was determined using Eq. \eqref{eq:BDF}, yielding a breakdown field (BDF) of 5.735 $\pm$ 0.2 MV/cm. The BDF reported in the literature for thin (ALD) hafnia is $\approx$10 MV/cm\cite{Saba2021}. The low BDF value measured here for the hafnia layer prevents the MOS structures from reaching deep accumulation.

The non-uniformity observed in the hafnia layer, along with its low relative permittivity and BDF, suggests the presence of nanopores and a low density relative to bulk hafnia. Referring to the fabrication flow, we note that the metal stack of the nanoantennas, Fig. \ref{fig:fab_flow}(D), consists of a layer of Cr, followed by a layer of Au, capped by a thin layer of Cr, implying that the hafnia layer must nucleate on this surface. We suspect an insufficient surface density of hydroxyl (OH) functional groups on this metal stack, compromising hafnia nucleation during the initial cycles of the ALD process and resulting in reduced density of the film\cite{Lemaire2017,Cremers2019}. Further work on improving the formation of hafnia on such metal stacks would be required.

\subsection{Epsilon near zero in ITO}

The deposited ITO layer exhibits an epsilon-near-zero (ENZ) state in the near-infrared (NIR) wavelength range, as confirmed by the analysis on witness samples (refer to Section S\ref{sup-section:ITOandHafniaCha} in the supplementary material for details). The device's performance is significantly influenced by the ENZ wavelength of the ITO layer. This perturbation in ITO permittivity near the nanoantenna, induced by carrier refraction, is facilitated by voltage gating, allowing for carrier density modulation through accumulation and depletion. The increased carrier density leads to a blue-shift in $\lambda_{ENZ}$ during accumulation impacts the reflectarray's reflection properties.

The wavelength  at which the refractive index of ITO becomes zero, $\lambda_{\text{ENZ}}$, can be determined using the following equation:

\begin{equation} \label{eq:enz_def}
\lambda_{\text{ENZ}} = \frac{2 \pi c_0 \sqrt{\epsilon_\infty}}{\omega_{D,b}}
\end{equation}

\noindent where $c_0$ is the speed of light in vacuum, $\epsilon_\infty$ denotes the dielectric constant of ITO, and $\omega_{D,b}$ corresponds to the Drude frequency for the bulk. Eq. \eqref{eq:enz_def} is derived from the Drude dispersion equation assuming that $\gamma_D \gg \omega_{D,b}$, as shown in Eq. \eqref{Sequ:DrudeModel} in the supplementary information file.

To estimate $\lambda_{\text{ENZ}}$ of the ITO layer in the reflectarray, we used the parameters obtained in Section \ref{sup-section:ITOandHafniaCha} of the supplementary information file. Given that the carrier concentrations measured via the Hall effect align well with $\omega_{D,b}$ derived from the ellipsometry fittings, we calculated $\lambda_{\text{ENZ}}$ using an average $\omega_D' = (\omega_{D\text{,Hall}} + \omega_{D\text{,Ell}})/2$, where $\omega_{D\text{,Hall}}$ is the Drude frequency for the bulk calculated using Eq. \eqref{Sequ:ncal} based on the carrier concentration measured via the Hall effect, and $\omega_{D\text{,Ell}}$ represents the Drude frequency for the bulk obtained from the ellipsometry fitting. Using Eq. \eqref{eq:enz_def}, the anticipated range of $\lambda_{\text{ENZ}}$ is from $\lambda_{\text{ENZ}} = 1.91$ $\mu$m to $\lambda_{\text{ENZ}} = 2.1$ $\mu$m. As the MOS structures enter the accumulation region, the carrier density in the perturbed ITO layer increases (Fig. \ref{fig:calculations1}(B)), resulting in a blue-shift of $\lambda_{\text{ENZ}}$ therein. If the breakdown field of hafnia is sufficiently high, $\lambda_{\text{ENZ}}$ may transition through the resonance wavelength of the reflectarray, leading to a significant phase shift in the reflected beam \cite{Calalesina2021}. Due to the low BDF of the hafnia layer here, it was not possible to apply a sufficiently large bias voltage for this regime to be attained.

\subsection{Speed of operation}
The response speed of the reflectarray is inherently limited by the time required for the MOS capacitors to reach full charge under accumulation, then to discharge when the bias voltage is reduced. The measured specific capacitances given in Fig. \ref{fig:calculations1}(A), the corresponding areas (\emph{cf.} Section S\ref{sup-section:pinAreas} in the supplementary information file), and the minimum and maximum pin resistances determined to be $R_{\text{pin}}=1.8$ k$\Omega$ and $R_{\text{pin}}=1.85$ k$\Omega$ (\emph{cf.} Section \ref{sup-section:pinResistance} in the supplementary information file), were considered in estimating the speed of operation. Based on these parameters, the time constant of the capacitors was calculated as $\tau = R_{\text{pin}}C_{\text{pin}}$, resulting in values ranging from $\tau=$ 1.98 to 2.5 ns, from which the maximum frequency of operation is determined to be 63 MHz (for $\tau=$ 2.5 ns).

\section{Conclusion}
We have successfully fabricated, experimentally characterized, and theoretically modelled a novel sub-wavelength reflectarray capable of varying the phase of the reflected beam without modifying its intensity or introducing scattering effects. 
The reflectarray pixels consisted of Au dipole nanoantennas integrated within a MOS capacitor structure. The pixels were fabricated on a fused Silica substrate, with a conformal coating of thin hafnia followed by layers of ITO and a Au backside mirror. By utilizing this MOS capacitor, we were able to modulate the carrier density of ITO at the interface with hafnia, thereby leading to the observed phase shift in the reflected beam. The reflectarray demonstrates excellent performance near 1550 nm, producing exclusively the desired specular reflected lobe. The reflectarray generated a maximum measured phase shift of about 30 degrees at $\lambda=1560$ nm, limited by the quality of our hafnia insulating film. The magnitude of reflection coefficient was about 0.5 and remained flat over the tuning range of the phase. 

\noindent\textbf{Research funding:}
Financial support provided by the Natural Sciences and Engineering Research Council of Canada (211636) and by Huawei Canada (570551) are gratefully acknowledged.

\noindent\textbf{Author contributions:} Luis Angel Mayoral Astorga obtained and processed the experimental data. Masoud Shabaninezhad performed all simulations. Howard Northfield, Ewa Lisicka-Skrzek, Sabaa Rashid and Spyridon Ntais designed layouts and fabricated the structures. Hamid Mehrvar, Eric Bernier, Dominic Goodwill, Lora Ramunno and Pierre Berini conceived and directed the research. All authors contributed to the discussion of the results and preparation of the manuscript. 

\noindent\textbf{Conflict of interest:} The authors declare no conflicts of
interests.

\noindent\textbf{Data availability:} All data generated or analyzed during this study are included in this published article and its supplementary information file.


\begin{thebibliography}{9}

\bibitem{Shalaginov2020}
Mikhail Y. Shalaginov, Sawyer D. Campbell, Sensong An et al., "Design for quality: reconfigurable flat optics based on active metasurfaces", \emph{Nanophotonics}, vol. 9, no. 11, 2020.

\bibitem{Nemati2018}
Arash Nemati, Qian Wang, Minghui Hong and Jinghua Teng, "Tunable and reconfigurable metasurfaces and metadevices", \emph{Opto‐Electron Adv.}, vol. 1, no. 5, 2018.

\bibitem{Shaltout2019}
Amr M. Shaltout, Vladimir M. Shalaev and Mark L. Brongersma, "Spatiotemporal light control with active metasurfaces", \emph{Science}, vol. 364, no. 6441, 2019.


\bibitem{Kim2021}
Inki Kim, Renato Juliano Martins, Jaehyuck Jang et al.,"Nanophotonics for light detection and ranging technology", \emph{Nature Nanotechnoly}, vol. 16, no. 5, 2021.

\bibitem{Heck2017}
Martijn J.R. Heck, "Highly integrated optical phased arrays: photonic integrated circuits for optical beam shaping and beam steering", \emph{Nanophotonics}, vol. 6, no. 1, 2017.

\bibitem{Berini2022}
Pierre Berini, "Optical beam steering using tunable metasurfaces", \emph{ACS Photonics}, vol. 9, no. 7, 2022.

\bibitem{Calalesina2021}
Antonino Cal\'a Lesina, Dominic Goodwill, Eric Bernier, Lora Ramunno and Pierre Berini, "Tunable plasmonic metasurfaces for optical phased arrays", \emph{IEEE Journal of Selected Topics in Quantum Electronics}, vol. 27, no. 1, 2021.


\bibitem{Decker2013}
Manuel Decker, Christian Kremers, Alexander Minovich et al., "Electro-optical switching by liquid-crystal controlled metasurfaces", \emph{Optics Express}, vol. 21, no. 7, 2013.

\bibitem{Olivieri2015}
Anthony Olivieri, Chengkun Chen, Sa\'{a}d Hassan, Ewa Lisicka-Skrzek, R. Niall Tait and Pierre Berini, "Plasmonic nanostructured metal–oxide–semiconductor reflection modulators", \emph{Nano Letters}, vol. 15, no. 4m 2015.

\bibitem{Thyagarajan2017}
Krishnan Thyagarajan, Ruzan Sokhoyan, Leonardo Zornberg and Harry A. Atwater, "Millivolt modulation of plasmonic metasurface optical response via ionic conductance", \emph{Advanced Materials}, vol. 29, no. 31, 2017.


\bibitem{Huang2016}
Yao-Wei Huang, Ho Wai Howard Lee, Ruzan Sokhoyan et al., "Gate-Tunable Conducting Oxide Metasurfaces", \emph{Nano Letters}, vol. 16, no. 9, 2016.

\bibitem{park2017}
Junghyun Park, Ju-Hyung Kang, Soo Jin Kim, Xiaoge Liu, and Mark L. Brongersma, "Dynamic Reflection Phase and Polarization Control in Metasurfaces", \emph{Nano Letters}, vol. 17, no. 1, 2017.

\bibitem{Shirmanesh2018}
Ghazaleh Kafaie Shirmanesh, Ruzan Sokhoyan, Ragip A. Pala, and Harry A. Atwater, "Dual-Gated Active Metasurface at 1550 nm with Wide (>300°) Phase Tunability", \emph{Nano Letters}, vol. 18, no. 5, 2018.



\bibitem{KimIl2020}
Sun Il Kim, Junghyun Park, Byung Gil Jeong et al., "Electrically Reconfigurable Active Metasurface for 3D Distance Ranging", 2020 IEEE International Electron Devices Meeting (IEDM), 2020.


\bibitem{KimIl2022}
Sun Il Kim, Junghyun Park, Byung Gil Jeong et al.,"Two-dimensional beam steering with tunable metasurface in infrared regime", \emph{Nanophotonics}, vol. 11, no. 11, 2022.

\bibitem{HowardN2021}
Howard Northfield, Oleksiy Krupin, R. Niall Tait and Pierre Berini, "Tri-layer contact photolithography process for high-resolution lift-off", \emph{Microelectronic Engineering}, vol. 241, no. 11545, 2021


\bibitem{Sabaa2021}
Sabaa Rashid, Jaspreet Walia, Howard Northfield et al., "Helium ion beam lithography and liftoff",\emph{Nano Futures}, vol. 5, no. 025003m, 2021.

\bibitem{CalaLesina2020}
Antonio Cala' Lesina, Dominic Goodwill, Eric Bernier, Lora Ramunno and Pierre Berini, "On the performance of optical phased array technology for beam steering: effect of pixel limitations", \emph{Opt. Express}, vol. 28, no. 21, 2020.

\bibitem{Araujo2009}
Marcos A. C. de Ara\'ujo, Rubens Silva, Emerson de Lima, Daniel P. Pereira
and Paulo C. de Oliveira, "Measurement of Gaussian laser beam radius using the knife-edge technique: improvement on data analysis", \emph{Applied Optics}, vol. 48, no. 2, 2009.

\bibitem{Innem2020}
Innem V. A. K. Reddy, Josep M. Jornet, Alexander Baev and Paras N. Prasad, "Extreme local field enhancement by hybrid epsilon-near-zero–plasmon mode in thin films of transparent conductive oxides", \emph{Optics Letters}, vol. 45, no. 20, 2020.


\bibitem{Rajput2023}
Swati Rajput, Vishal Kaushik, Lalit Singh et al., "Efficient optical modulation in ring structure based on Silicon-ITO heterojunction with low voltage and high extinction ratio", \emph{Optics Communication}, vol. 545, no. 129562, 2023.

\bibitem{Luscombe1992}
James H. Luscombe, Ann M. Bouchard and Marshall Luban, "Electron confinement in quantum nanostructures: Self-consistent poisson-Schr\"odinger theory", \emph{Phys. Rev. B}, vol. 46, no. 10262, 1992.

\bibitem{Fiegna1997}
C. Fiegna and A. Abramo, "Solution of 1-D Schrodinger and Poisson equations double gate SOI MOS" ,SISPAD '97. 1997 international conference on simulation of semiconductor processes and devices. Technical digest, 1997.

\bibitem{Sojib2022}
Mohammad Sojib, Dhruv Fomra, Vitaliy Avrutin, \"{U}. \"{O}zg\"{u}r and Nathaniel Kinsey, "Optimizing epsilon-near-zero based plasmon assisted modulators through surface-to-volume ratio", \emph{Optics Express}, vol. 30, no. 11, 2022.

\bibitem{Masoud2022}
Masoud Shabaninezhad, Lora Ramunno and Pierre Berini, "Tunable plasmonics on epsilon-near-zero materials: the case for a quantum carrier model", \emph{Optics Express}, vol. 30, no. 26, 2022.

\bibitem{comsol6v1}
COMSOL Multiphysics\copyright V 6.1, "COMSOL AB, Stockholm, Sweden", www.comsol.com.

\bibitem{Kittel2004}
Charles Kittel, \emph{Introduction to solid state physics. Free electron fermi gas.}, Wiley, 2004.


\bibitem{Berini2000}
Pierre Berini, "Plasmon-polariton waves guided by thin lossy metal films of finite width: Bound modes of symmetric structures", \emph{Physical Review B}, vol. 61, no. 10484, 2000.

\bibitem{Johnson1972}
P. B. Johnson and R. W. Christy, "Optical constants of the noble metals", \emph{Phys. Rev. B}, vol. 6, no. 12, 1972.

\bibitem{Rakic98}
Aleksandar D. Raki\'{c}, Aleksandra B. Djuri\v{s}i\'{c}, Jovan M. Elazar and Marian L. Majewski, "Optical properties of metallic films for vertical-cavity optoelectronic devices", \emph{Appl. Opt.}, vol. 37, no. 22, 1998.

\bibitem{Lee2013}
Chien-Wei Lee and Jenn-Gwo Hwu, "Quantum-mechanical calculation of carrier distribution in MOS accumulation and strong inversion layers", \emph{AIP Advances}, vol. 3, no. 10, 2013.

\bibitem{Banalis1989}
Constantine A. Balanis, "Geometrical theory of diffraction: edge diffraction" in \emph{Advanced engineering electromagnetics}, Wiley press, 1989.

\bibitem{Kenta2019}
Kenta Suzuki, Kenichi Oguchi, Yasuaki Monnai, Makoto Okano and Shinichi Watanabe, "Spatio-temporal imaging of terahertz electric-field vectors: observation of polarization-dependent knife-edge diffraction", \emph{Applied Physics Express}, vol. 12, no. 052010, 2019.


\bibitem{Alshehab2019}
Maryam Al-Shehab, Saba Siadat Mousavi, Maude Amyot-Bourgeois et al., "Design and construction of a Raman microscope and characterization of plasmon-enhanced Raman scattering in graphene", \emph{Journal of the Optical Society of America B}, vol. 36, no. 8, 2019.

\bibitem{Saba2021}
Saba Siadat Mousavi, Anthony Olivieri and Pierre Berini, "Fabrication of a high-speed plasmonic reflection/transmission modulator", \emph{AIP Advances}, vol. 11, no. 025023, 2021.

\bibitem{Lemaire2017}
Paul C. Lemaire, Mariah King and Gregory N. Parsons, "Understanding inherent substrate selectivity during atomic layer deposition: Effect of surface preparation, hydroxyl density, and metal oxide composition on nucleation mechanisms during tungsten ALD", \emph{J. Chem. Phys.}, vol. 146, no. 052811, 2017.

\bibitem{Cremers2019}
V\'eronique Cremers, Riikka L. Puurunen, Jolien Dendooven, "Conformality in atomic layer deposition: Current status overview of analysis and modelling", \emph{Appl. Phys. Rev.}, vol. 6, no. 021302, 2019.


\end{thebibliography}
\end{document}


\title{Electrically tunable plasmonic metasurface as a matrix of nanoantennas: supplemental material}
 \maketitle

\section{ITO and hafnia characterization}
\label{sup-section:ITOandHafniaCha}

During the deposition process of the indium tin oxide (ITO) and hafnia layers, witness samples were created with the purpose of evaluate the hafnia and ITO layers. Sample SR83 had a target thickness of 79 nm of ITO deposited on fused Silica. Sample Hf4-227 consisted of a target thickness of 7 nm of hafnia deposited on fused Silica. Additionally, a bilayer sample (Hf3-227) was created, which consisted of a target thickness of 79 nm of ITO on top of 7 nm hafnia, both deposited on a fused Silica substrate.

The electronic properties of the ITO layer were analyzed using an Hall effect measurement system (HMS-3000, Ecopia) in the Van der Pauw configuration. The system measured the resistivity, carrier concentration, and mobility \cite{Hemenger1973}. The Spring Clip Board Sample Mounting Boards (SPCB1, Ecopia) was used to hold and probe the sample, which had to be cut into squares of dimensions approximately 20 mm $\times$ 20 mm. Conductive ink drops were placed on the corners of the cut samples to ensure good electrical contact between the ITO surface and the needle probes. A magnetic field from a permanent magnet with a magnetic flux density of 0.55 T was used during the Hall effect measurements.

The optical characteristics of the deposited ITO and hafnia layers were extracted by taking ellipsometry measurements using a Spectroscopic Ellipsometer (UVISEL Plus, Horiba Scientific). This system uses a double modulation acquisition technology to measure the ellipsometry angles $\Psi$ and $\Delta$ in reflection. The angle of incidence was kept fixed at 70$^{\text{o}}$. The collected ellipsometry data were fitted to a layered model to determine the optical parameters of the ITO and hafnia layers. The algorithm used for the fitting was the Levenberg-Marquardt algorithm. The goodness of  fit (GOF) was determined by:
 \begin{equation}
  \chi^2 = \text{min} \sum_{i=1}^n\Big[ \frac{(\Psi_{th}-\Psi_{exp})^2}{\Gamma_\Psi}  +  \frac{(\Delta_{th}-\Delta_{exp})^2}{\Gamma_\Delta} \Big ]
 \end{equation}
where ($\Psi_{th}$, $\Delta_{th}$) and ($\Psi_{exp}$, $\Delta_{exp}$) are the theoretical and measured ellipsometry angles, and $\Gamma_{x}$ is the standard deviation of each data point.
 
The refractive index of hafnia was modelled using the Cauchy dispersion model, 
 $n(\lambda)=A+B\lambda^{-2}+C\lambda^{-4}$; $k=0$. ITO was modelled using Drude dispersion model:
 
 \begin{equation} \label{Sequ:DrudeModel}
\epsilon(\omega)=\epsilon_\infty - \frac{\omega_{D,b}^2}{\omega^2+i\gamma_D\omega} .  
 \end{equation}

To analyze the transparent samples, we employed a backside reflection model. We used void as the substrate and added a 500 $\mu$m fused Silica layer. The total number of backside reflected beams was kept constant at 2. The percentage of first collected beam was set at 100\% and the last collected beam was adjusted during the fitting process in order to minimize the GOF. The fused Silica layer was modeled via a multi-Sellmeier model, utilizing the bulk parameters reported in \cite{Tan1998}.

The ellipsometry fitting process began by extracting the optical parameters of hafnia from the Hf4-227 sample. Next, the Drude parameters for ITO were determined from sample SR83. Subsequently, the hafnia parameters obtained from the Hf4-227 sample were used to model the hafnia layer on sample Hf3-227. For the ITO layer on the Hf3-227 sample, we used the Drude parameters obtained from the SR83 analysis as initial values in the fitting process.

To fit the ellipsometry data measured from Hf4-227, we used initial parameters for the Hafnia layer that correspond to bulk hafnia \cite{Kuhaili2004}. Specifically, $A=1.875$, $B=6.28\times 10^3$ nm$^2$ and $C=5.8\times 10^8$ nm$^4$ in the Cauchy dispersion model. The thickness of the hafnia layer and the parameter A of the Cauchy dispersion model were used as fitting parameters, and the fitting wavelength range was set from 1000 to 2000 nm. We also included a roughness of 1 nm in the model, accounting for 50\% air and 50\% hafnia.
 
During the fitting for the Drude parameters of the ITO on sample SR83, we used initial parameter values for ITO as the average of values reported in \cite{Zhaolin2012, Cleary2018, Masoud2022}: $\epsilon_\infty=4$, $\omega_{D,b}=2\times 10^{15}$ rad/sec, and $\gamma_D=1.3\times 10^{13}$ rad/sec. The thickness of the ITO layer was also a fitting parameter with initial value of 79 nm.

The model for the sample Hf3-227 consisted of a top layer of ITO over hafnia. The thickness of the ITO and the Drude parameters of the ITO layer were used as fitting parameters. The Hafnia layer was fixed at 7.2 nm which resulted from the fitting process on the sample Hf3-227. The initial thickness of the ITO layer was 79 nm and the initial parameters for the ITO were the resulted parametes from the fitting performend on sample SR83.

The resulted fitting process using the ellipsometry data from the witness samples were consistently very good. The GOF obtained was low across all cases studied. Specifically, for the model of sample Hf4-227 (hafnia on fused Silica), the value of $\chi^2$ was 0.065, yielding a thickness of hafnia of 7.196 $\pm$ 0.862 nm and an $A$ value of 1.872126 $\pm$ 0.0379. Similarly, for the model of sample SR83, the value of $\chi^2$ was 0.074, and the resulting Drude parameters were $\epsilon_\infty=4.0163 \pm 0.02$, $\omega_{D,b}=1.7521 \times 10^{15} \pm 0.0105 \times 10^{15}$ rad/sec, and $\gamma_D=0.1304 \times 10^{15} \pm 0.0032\times 10^{15}$ rad/sec, with a ITO thickness of 79.16 $\pm$ 0.97 nm. Finally, for the model of sample hf3-227 (ITO on 7.2 nm hafnia on fused Silica), the value of $\chi^2$ was 0.5, yielding an ITO thickness of 70 $\pm$ 2.6 nm and Drude parameters $\epsilon_\infty=4.026 \pm 0.04152$, $\omega_{D,b}=1.9515 \times 10^{15} \pm 0.015 \times 10^{15}$ rad/sec, and $\gamma_D=0.1308 \times 10^{15} \pm 0.004 \times 10^{15}$ rad/sec.

The Hall effect measurements conducted on samples SR83 and Hf3-227 yielded the following results: for sample SR83, a carrier concentration of $n_b=-3.32\times10^{20}$ cm$^{-3}$, resistivity of $\rho=0.428\times10^{-3}$ $\Omega$cm, and mobility of $\mu=43.93$ cm$^{2}$/Vs. For sample Hf3-227, a carrier concentration of $n_b=-4.433\times10^{20}$ cm$^{-3}$, resistivity of $\rho=0.337\times10^{-3}$ $\Omega$cm, and mobility of $\mu=38.24$ cm$^{2}$/Vs were obtained.

The measured carrier concentration was in good agreement with the predicted Drude frequency $\omega_{D,b}$. This was confirmed by using 
\begin{equation} \label{Sequ:ncal}
|n_{b,ell}|= \frac{\omega_{D,b}^2 \epsilon_0 m_n^*}{e^2 \times 10^6} 
\end{equation}
where $|n_{b,ell}|$ is the carrier concentration in cm$^{-3}$, $\omega_{D,b}$ is the Drude frequency in rad/sec obtained from the fitting process, $\epsilon_0$ is the permittivity of vacuum in F/m, $m_n^*=0.35 m_e$ is the effective electron mass in Kg, $m_e$ is the electron mass in Kg, and $e$ is the electron charge in Coulombs.

The measured carrier concentration and the estimated value obtained through ellipsometry fitting showed a good agreement per sample. For sample SR83, the ellipsometry-derived carrier concentration $|n_{b,ell}|$ is 3.37 $\times 10^{20}$, which differs by 1.5\% from the carrier concentration measured using the Hall effect system. Similarly, for sample Hf3-227, the ellipsometry-derived carrier concentration $|n_{b,ell}|$ is 4.19 $\times 10^{20}$, exhibiting a difference of 5.5\% compared to the carrier concentration measured using the Hall effect system.

The relative density of the hafnia layer can be estimated using the Lorentz-Lorenz (LL) equation. The relationship between the variation of the film density $\rho_f$ and the refractive index of the fitted hafnia layer is described by the equation $\rho_f/\rho_{145}=(n_f^2-1)(n_{145}^2+2)/[(n_f^2+2)(n_{145}^2-1)]$, where $\rho_{145}$ is the density and $n_{145}$ is the refractive index of 145 nm thick hafnia reported in \cite{Kuhaili2004}. The LL equation is valid for a transparent film. Using $n_{145}=1.878$ and $n_f=1.851$ at $\lambda=1.56$ $\mu$m, the relative density of the hafnia layer is $\rho_f/\rho_{145}=0.97$ which indicates a good dense hafnia layer on the witness samples.

\section{Illustrative Case of Phase Shift Measurement}
\label{sup-section:phase_measurement_example}

\begin{figure}[ht!]
\centering
\includegraphics[width=0.6\textwidth]{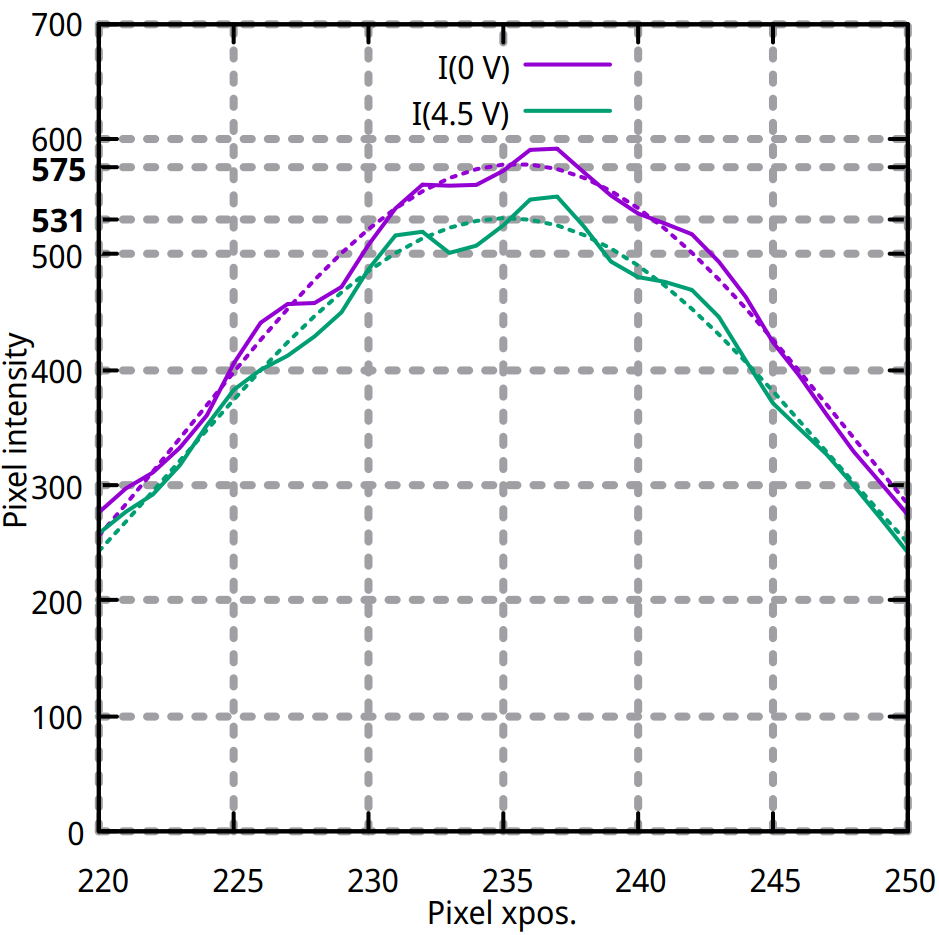}
\caption{Intensity of the interference pattern measured at 0 V and 4.5 V. The solid lines represent the raw measurement data, while the dashed lines corresponds to a Gaussian fittings.}
 \label{figS:Intensity_showcase}
\end{figure}

Figure \ref{figS:Intensity_showcase} displays the intensity of the interference pattern measured at 0 V and 4.5 V at 1560 nm wavelength, along with their corresponding Gaussian fittings. The intensity is presented in the digital scale of the IR camera. The peak values of the Gaussian fittings were utilized for calculating the phase shift. Initially, the value $A=\frac{\sqrt{I(0 V)}}{2}=287.5$ is computed. Subsequently, using the value of $I(4.5 V)=531$ in the phase shift equation $\phi(V)=\cos^{-1}\Big({\frac{\sqrt{I(V)}-A}{A} }\Big)$:

\begin{equation}
\begin{split}
\phi(4.5 V)=\cos^{-1}\Big({\frac{\sqrt{I(4.5)}-A}{A} }\Big)\\
\phi(4.5 V)=\cos^{-1}\Big({\frac{\sqrt{531}-287.5}{287.5} }\Big)\\
\phi(4.5 V)=22.8 ^\circ
\end{split}
\end{equation}

\section{Raw data measurements of the capacitance and consideration of the instrument accuracy}
\label{sup-section:rawdata}
\begin{figure}[ht!]
\centering
\includegraphics[width=0.6\textwidth]{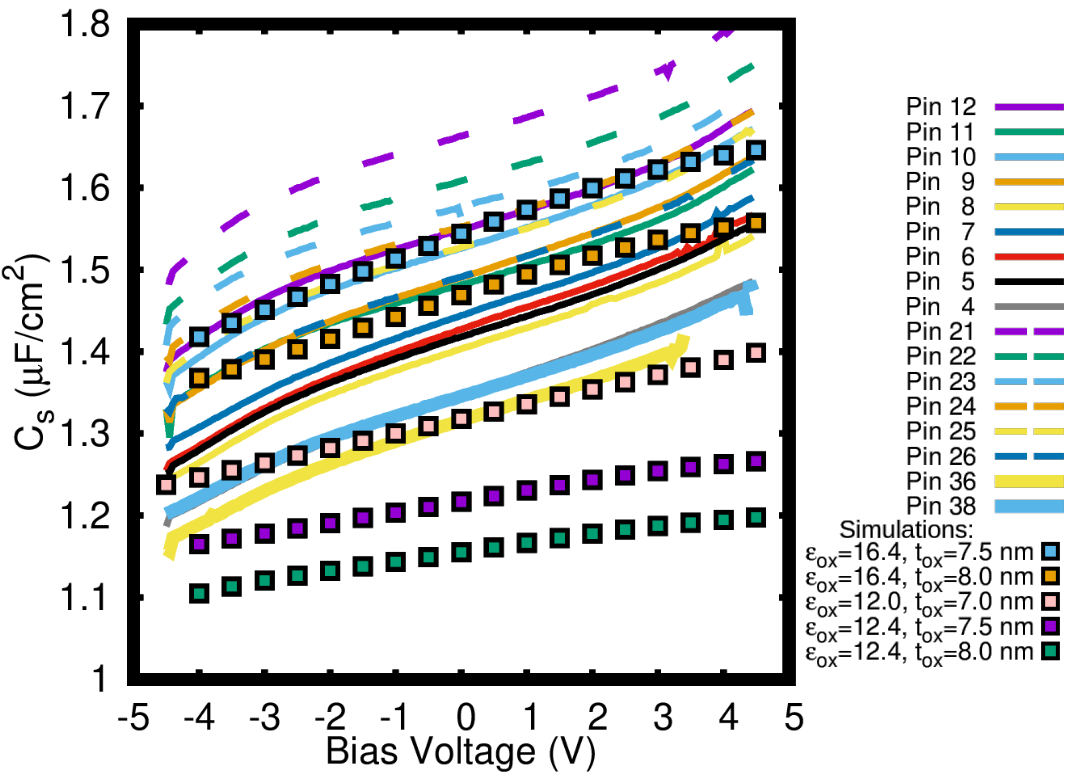}
\caption{Experimental capacitance per unir area of sub-wavelength reflectarray with parasitic capacitance offset of $\approx 0.1$ $\mu F/cm^2$ along with calculated C-V characteristics.}
 \label{figS:rawdata}
\end{figure}

Table \ref{table:S_rawdata} displays the raw experimental measurements of capacitance per pin, including the corresponding offset from the zero value ($\Delta C$). The offsets presented are in close proximity to the accuracy limit of our LCR meter, specified to be around 0.05 pF. Figure \ref{figS:rawdata} illustrates the corresponding specific capacitance calculated using the reported pin area, detailed further in this document, alongside the theoretical calculation for comparition.

 \begin{table} [!ht] 
\centering
\caption{Raw capacitance measurements per pin.}
\begin{tabular}{cccc}
Pin & C at 4.5V [pF]	& $\Delta$C (pF) \\ \midrule
12 & 1.23692068047208 & 0.0730293896 \\
11 & 1.16947351301844 & 0.0720869324 \\
10 & 1.26052976060512 & 0.0754157948 \\
9 &	1.26908116244908 & 0.0775309684 \\
8  & 1.23367700351742 & 0.0800470418 \\
7 &	1.30277889079296 & 0.0820291584 \\
6 &	1.30803768792516 & 0.0834921226 \\
5 &	1.31317071641776 & 0.0844330742 \\
4 &	1.25982232351164 & 0.0847897002 \\
21 & 1.32253303390016 & 0.0730293896 \\
22 & 1.26111042148532 & 0.0720869324 \\
23 & 1.29350908767116 & 0.0754157948 \\
24 & 1.31274660385196 & 0.0775309684 \\
25 & 1.33764210140726 & 0.0800470418 \\
26 & 1.34026621618176 & 0.0820291584 \\
38 & 1.11178718852108 & 0.0754157948 \\
\end{tabular}
\label{table:S_rawdata}
\end{table}

\section{Pin serial resistance}
\label{sup-section:pinResistance}

\begin{figure}[ht!]
\centering
\includegraphics[width=0.6\textwidth]{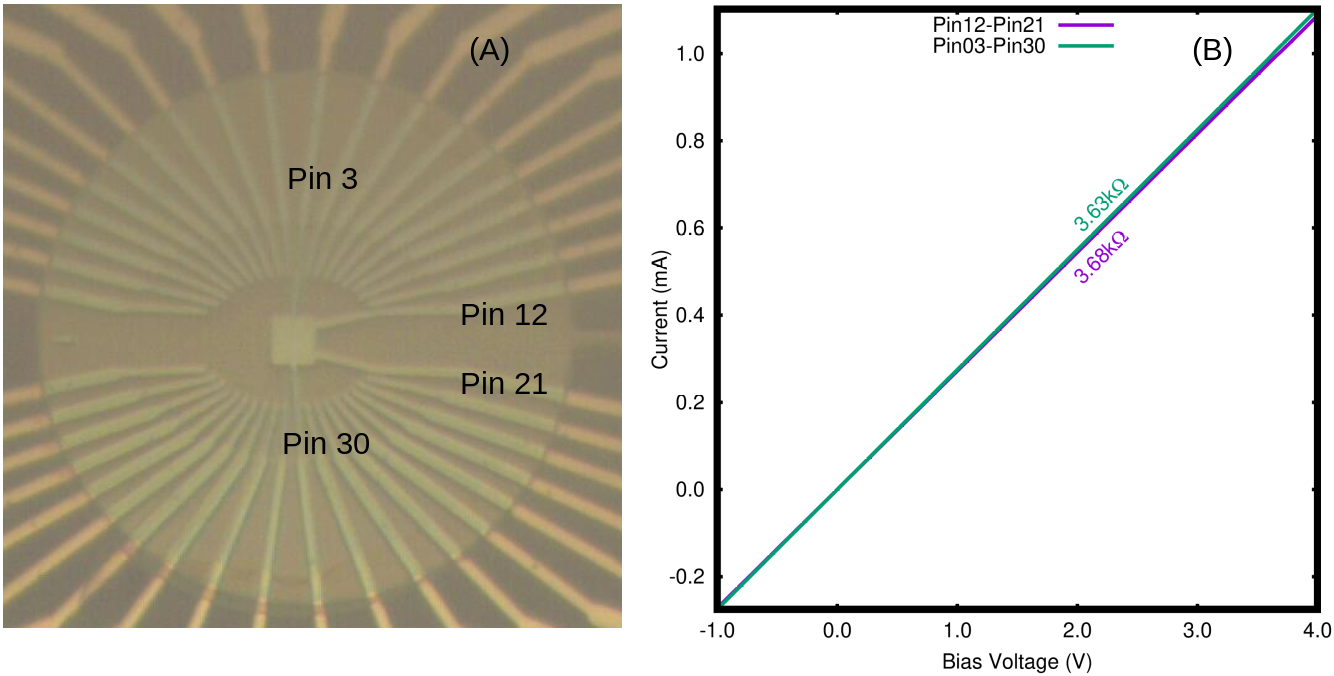}
\caption{(a)Test structure for DC testing. (b) Pin to pin current-voltage (I-V) characteristics.}
 \label{figS:testStructure}
\end{figure}

Test structures were defined on the first metal layer to study the pin and fan-out resistance. Fig. \ref{figS:testStructure}(A) shows the test structure after deposition of the hafnia layer. Notably, fan-out fingers and pins 12, 21, 3 and 30 are short-circuited together. It is important to mention that fingers 3 and 30 have the same dimensions (thickness, length, width), as do fingers 12 and 21, the former pair being of the shortest length and the latter pair being of the longest length.

The voltage-current characteristics of the pins were measured using a source-meter (2400, Keithley), which allows for voltage sourcing during measurements without the need for connection changes. A bias voltage was applied between pins 3 and 30 or between pins 12 and 21, and the resulting current was recorded. Fig. \ref{figS:testStructure}(B) displays the measured current for bias voltages ranging from -1 to 4 V. The I-V characteristics yield a consistent load 3.63 k$\Omega$ for the finger pair 3-30 and 3.68 k$\Omega$ for the finger pair 12-21. Based on these measurements, we determined that a single fan-out finger and pin resistance ranged from 1.81 to 1.84 k$\Omega$.

\section{Capacitor areas}
\label{sup-section:pinAreas}
\begin{figure}[ht!]
\centering
\includegraphics[width=0.6\textwidth]{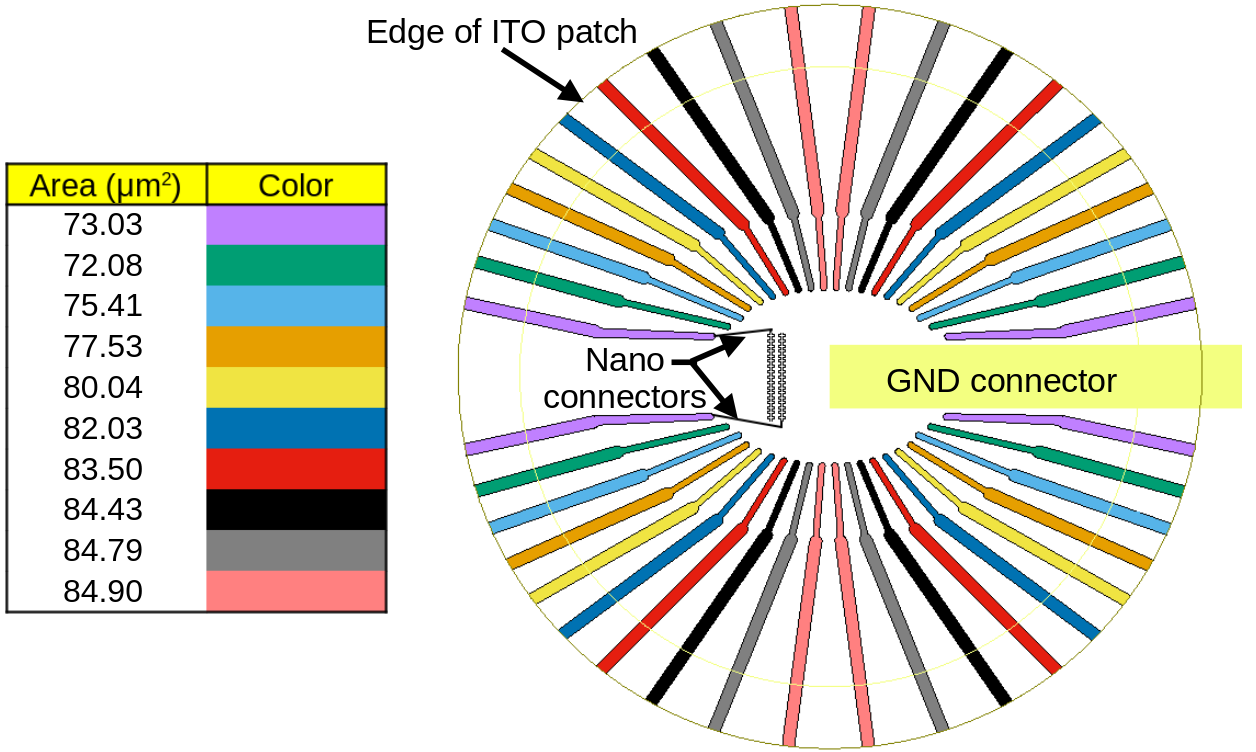}
\caption{Pin areas beneath the ITO patch.}
 \label{figS:pinArea}
\end{figure}

The area of each capacitor was taken as the area of each fan-out finger beneath the ITO patch, as shown in Fig. \ref{figS:pinArea}. The areas were calculated from the layout file of the first metal layer. Moreover, the area of the nano-connectors from the fingers to the nano-array and the area of the nano-antennas were considered. Specifically, the area of the nano-connectors were 0.731524 $\mu$m$^2$. The area of the nano-antennas was calculated as $(10$ $\mu$m$/a_y)[L_dW+(a_y-w)w_c]=0.81327$ $\mu$m$^2$, where the dimensions are in $\mu$m.

